\documentclass[preprint,12pt]{elsarticle}
\usepackage{graphicx}
\usepackage[margin=1.0in]{geometry}
\usepackage{color, colortbl}
\usepackage{hyperref}
\usepackage{float}
\usepackage{lineno}

\usepackage{subcaption}

\definecolor{LightCyan}{rgb}{0.88,1,1}
\definecolor{LightRose}{rgb}{1,0.88,0.88}
\definecolor{LightGreen}{rgb}{0.88,1,0.88}

\title{Convolutional Auto-Encoders for Drift Chamber de-noising for CLAS12}
\author[1]{Gagik Gavalian}
\author[2]{Polykarpos Thomadakis}
\author[2]{Angelos Angelopoulos}
\author[2]{Nikos Chrisochoides}

\address[1]{Jefferson Lab, Newport News, VA, USA}
\address[2]{CRTC, Department of Computer Science, Old Dominion University, Norfolk, VA, USA}

\begin{document}

\begin{abstract}
In this article, we present the results of using Convolutional Auto-Encoders for de-noising raw data for CLAS12 drift chambers.
The de-noising neural network provides increased efficiency in track reconstruction, also improved performance for high 
luminosity experimental data collection. The de-noising neural network used in conjunction with the previously developed track 
classifier neural network~\cite{Gavalian:2022hfa} lead to a significant track reconstruction efficiency increase for current luminosity
($0.6\times10^{35}~cm^{-2}~sec^{-1}$ ). The increase in experimentally measured quantities will allow running experiments at twice 
the luminosity with the same track reconstruction efficiency. This will lead to huge savings in accelerator operational costs, and large 
savings for Jefferson Lab and collaborating institutions.
\end{abstract}
\maketitle

\section{Introduction}

During the past few years, there was a big interest in using Artificial Intelligence (AI) in 
various areas of nuclear physics, from data processing to physics analysis. With continuously 
improving methods of Machine Learning (ML) and computational hardware, it becomes easy to 
substitute some computational tasks with ML algorithms leading to a smaller and computationally
more efficient codebase. In this article, we discuss the implementation of Convolutional Auto-Encoders 
for de-noising data from CLAS12~\cite{Burkert:2020akg} tracking detectors (Drift 
Chambers~\cite{Mestayer:2020saf}). The de-nosing was used to analyze simulated data to measure
improvement in track reconstruction efficiency.

\section{CLAS12 Drift Chambers}

The Drift Chambers (DC), which are  part of the large detector system of CLAS12 located in the experimental 
Hall-B at Jefferson Lab. They are used for charged particle detection in the forward direction 
(covering polar angles $5-35^\circ$). The CLAS12 forward detector is built around a six-coil 
toroidal magnet which divides the active detection area into six azimuthal regions, called “sectors”. 
For each sector, there are separate drift chambers installed consisting of 3 regions. Each region contains 
two super-layers, each of them containing 6 layers of wires.   Each layer of the drift chamber 
consists of 112 signal wires making each region a matrix of 12x112. The raw signal from 
one sector makes a matrix of 36x112, which is analyzed independently from other sectors
to extract trajectories of charged particles from raw signals. 

 \begin{figure}[!h]
\begin{center}
 \includegraphics[width=3in]{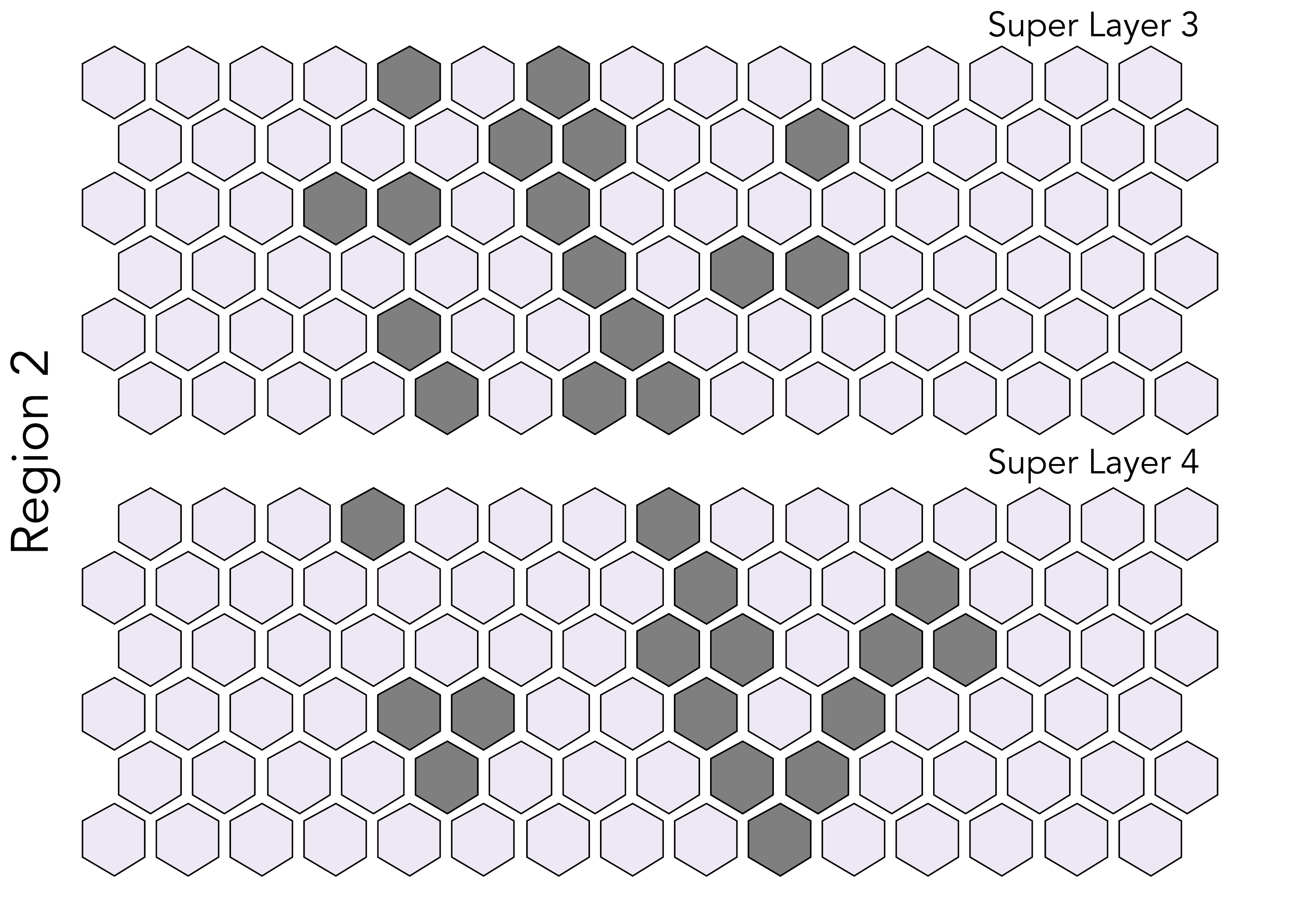}
 \includegraphics[width=3in]{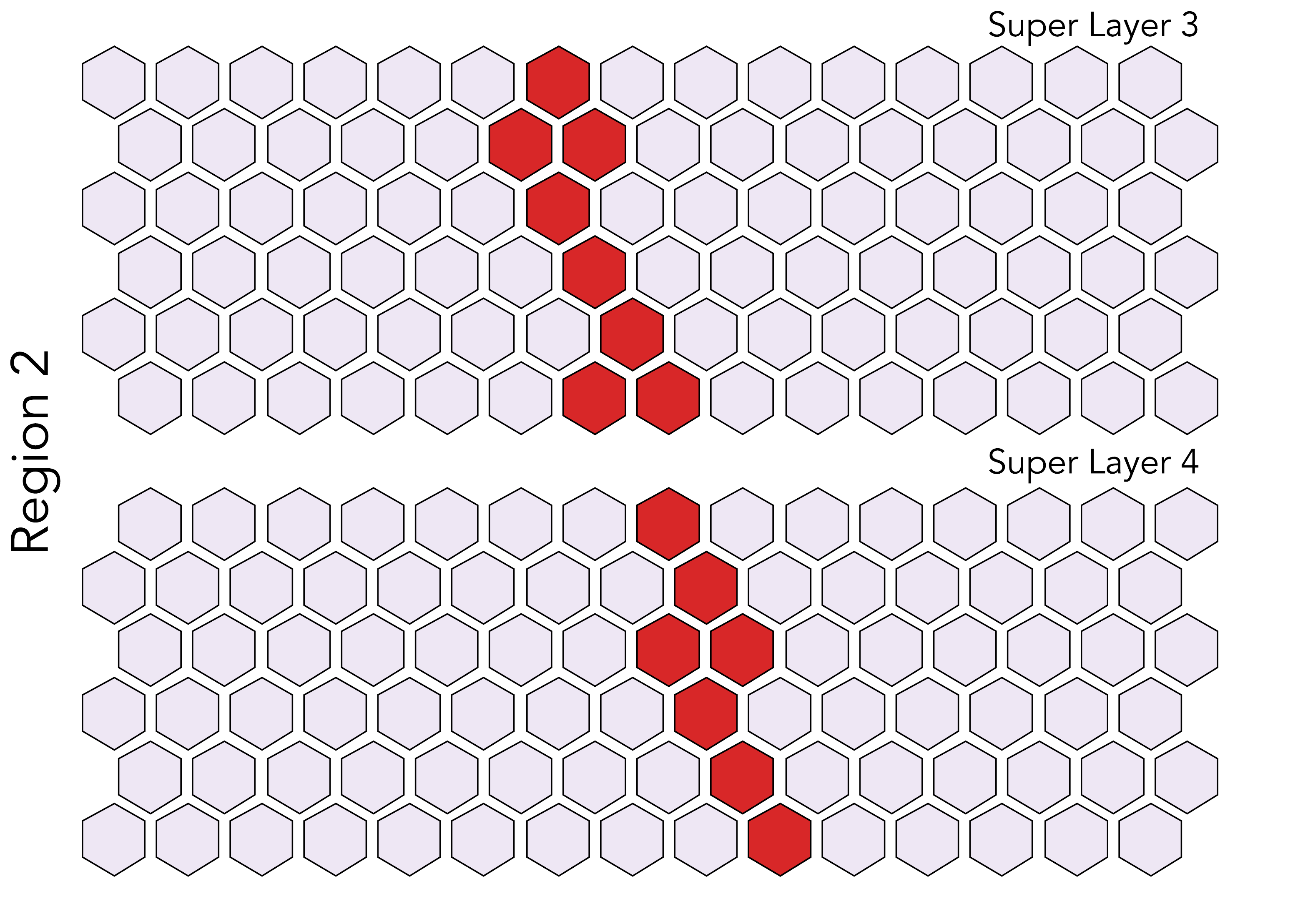}
\caption {Example of clustering for one region of Drift Chambers. The left panel shows 
all the hits detected in the drift chamber (for this particular region), and the right panel 
shows results of clustering where some hits were identified as a background and were removed,
and the remaining hits were grouped to form a cluster.}
 \label{conv:denoising}
 \end{center}
\end{figure}

\begin{figure}[!h]
\begin{center}
 \includegraphics[width=6.0in]{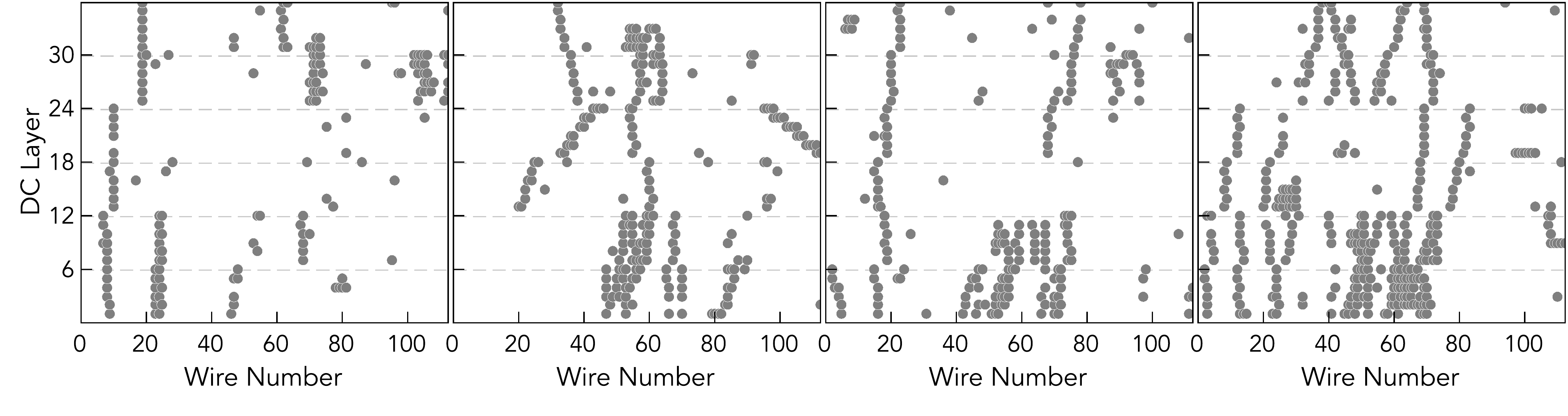}
 \includegraphics[width=6.0in]{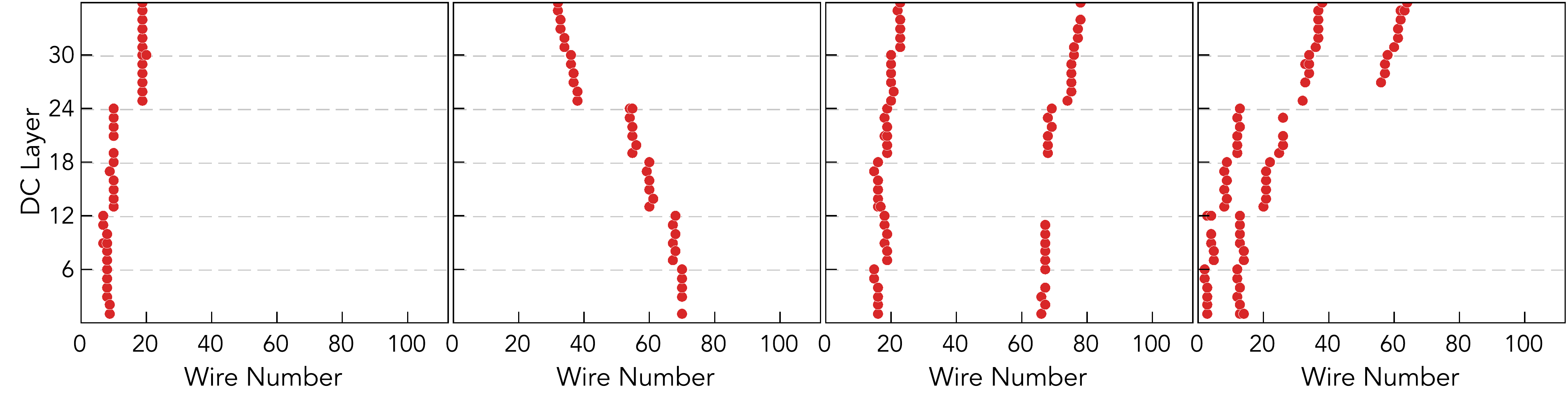}
\caption {Example of reconstructed tracks in drift chambers. The signal hits in drift chambers are shown 
on the top row. The hits (clusters) belonging to identified tracks are shown on the bottom row. Dashed 
lines represent the boundaries of super-layers.}
 \label{conv:trackfinding}
 \end{center}
\end{figure}

Each super-layer is analyzed separately for each sector and hits grouped together along the track trajectory 
are combined into clusters (or segments). In Figure~\ref{conv:denoising} the procedure is shown for one
region where all the hits (dark gray) are shown on the left panel, and clusters (red) are shown on the right panel,
by grouping neighboring wires after removing noise hits. Each super-layer
may have multiple clusters. The tracking algorithm creates a list of track candidates consisting of one cluster 
per super-layer and then analyzes the list to determine which candidates form a valid track. The identified 
tracks are further refined by passing them through Kalman filter~\cite{Kalman1960}. 
Examples of analyzed events in one sector can be seen in Figure~\ref{conv:trackfinding}, where 36x112 matrices
for four sectors are shown (not from the same event) with all signal hits in all layers (top row). The hits 
for clusters for identified tracks are shown on the bottom row.

As can be seen from the figure one or multiple tracks can be detected in one sector for the event. The efficiency
of finding these tracks depends on the cluster finding algorithm. With increased luminosity, the number of background 
hits increases, and it becomes difficult to separate background hits from signal hits due to heavy overlap between them.
This results in lost clusters and eventually in a decrease in track finding efficiency. 
In this work, Machine Learning is used to remove background hits prior to the clustering algorithm to improve cluster 
finding and consequentially track finding efficiency. The reconstructed experimental data is used to train Convolutional
Auto-Encoder for de-noising the drift chamber signal~\cite{Thomadakis:2022zcd}.

\section{Neural Network}

\begin{figure}[!h]
\begin{center}
 \includegraphics[width=5.1in]{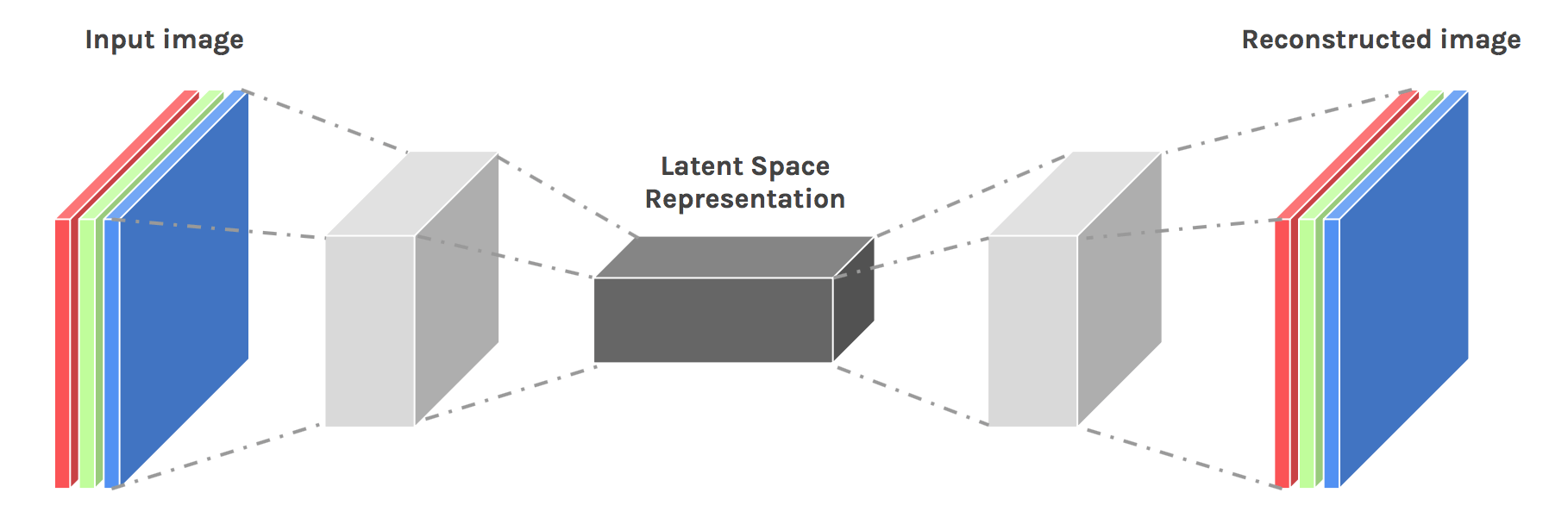}
\caption {De-noising Convolutional Auto-Encoder architecture. }
 \label{network:cnn_encoder}
 \end{center}
\end{figure}

The Convolutional Auto-Encoder is used to de-noise raw data from the CLAS12 drift chambers~\cite{Thomadakis:2022zcd}. 
The input and output for the network are matrices of size 36x112 representing hits in one sector of drift chambers. 
The training data was extracted from experimental data processed with CLAS12 reconstruction software. 
The raw hits (converted into a matrix) are used as an input for the neural network and a matrix constructed 
only from hits that belong to reconstructed tracks as an output (see Figure~\ref{conv:trackfinding}). 
In the training data set multiple track hits were allowed in the output matrix, shown on Figure~\ref{conv:trackfinding}.
The structure of the neural network can be seen in Figure~\ref{network:cnn_encoder}, where the input and the output are images
of size 36x112. Convolutional and Max Pool layers are used for encoding the image into smaller latent space and then
decoding it into an output image (of the same size as input) that contains only the desired pixels activated.

\begin{figure}[!h]
\begin{center}
 \includegraphics[width=5.8in]{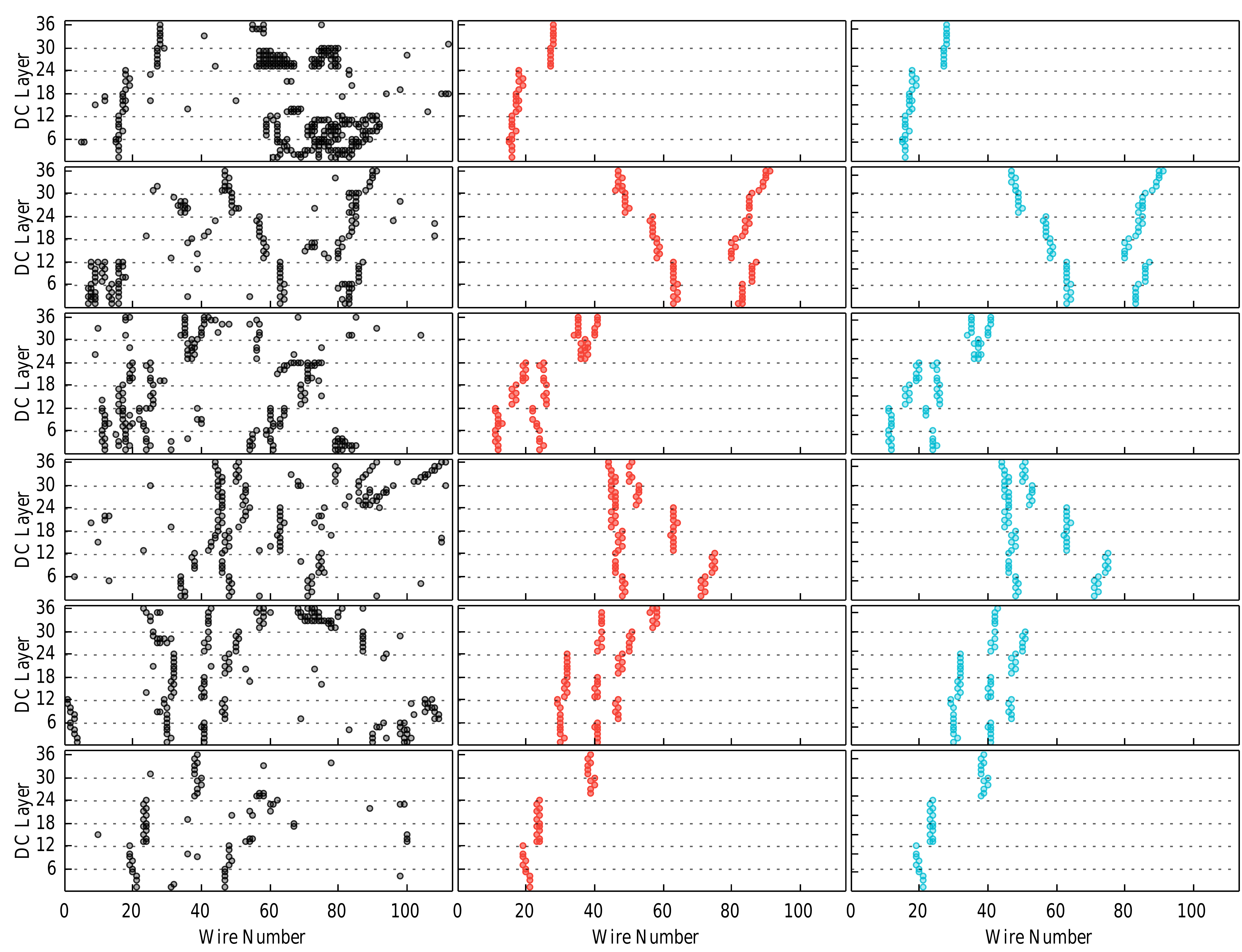}
\caption {Results from the de-noising auto-encoder. The raw hits are shown
in the left column for five random events, along with hits reconstructed by the 
CLAS12 tracking algorithm in the middle column. The resulting  hits matrix 
from the de-noising raw hits are shown in the right column. (Systematic studies 
of de-noiser performance can be found here~\cite{Thomadakis:2022zcd})}
 \label{network:cnn_results}
 \end{center}
\end{figure}

The networks are validated on experimental data where the number of hits along the 
track trajectory from de-noised are compared to the hits reconstructed by the conventional algorithm
as part of a valid track. An example of comparison can be seen in Figure~\ref{network:cnn_results} 
where raw data (left column) are shown along with data with hits belonging to reconstructed tracks 
identified by the conventional tracking algorithm (middle column) and 
reduced data processed by a de-noising neural network (right column).

As can be seen from the figure, the de-noising neural network removes all background hits not 
associated with a track, while preserving hits belonging to a track. 
Systematic studies~\cite{Thomadakis:2022zcd} showed that more than 
$95\%$ of the track related hits are preserved in the output of denoiser while 
background hits are significantly suppressed for normal experimental conditions of $45~nA$
incident beam current. Systematic studies showed that in more than $85\%$ of cases all 
6 clusters belonging to the track are fully identified by the algorithm after de-noising, 
and in more than $97\%$ of the cases, 5 clusters from the original track are recovered. The CLAS12 
track reconstruction algorithm can reconstruct tracks with 6 or 5 clusters along the 
track trajectory, which means that even if some clusters are lost due to de-noising 
procedure the track efficiency does not suffer significantly from this.

For our implementation of de-noising software, we used TensorFlow/Keras~\cite{keras-website} to train 
and evaluate the network. The resulting network parameters (weights) were saved 
in a HDF5 file. The denoiser implementation for the CLAS12 reconstruction software is 
done using DeepLearning4J~\cite{dl4j-website} which supports model imports 
through HDF5 files. 
The data analysis and data visualization are done using the GROOT~\cite{groot-github} visualization 
package, developed for the CLAS12 software infrastructure (in Java) and is publicly available 
through github releases. GROOT is also included in Jas4pp\cite{Chekanov:2020bja}  
(data-analysis framework for physics and detector studies). The de-noising is not yet implemented 
as a part of the CLAS12 reconstruction workflow and works as a standalone package to process raw data before 
they are analyzed with reconstruction software.

\section{Data Description}

\subsection{Monte-Carlo Simulation}

For these studies, we used physics reactions generated using Pythia Monte-Carlo~\cite{Pythia:2022} event generator, and generated events were processed with GEMC~\cite{gemc:2022} (GEANT4~\cite{geant4:2022} based detector simulation program) to produce data similar to experimental data. The four charged particle final state (namely $e^-,\pi^+,\pi^-,p$) is selected in the output of Pythia for our studies.
In addition, any number of neutral particles is allowed.

Simulated physics events were processed with the CLAS12 emulation software (GEMC) that produces raw signal data (similar to experimental).
Using these generated files new files were generated emulating different luminosity experimental conditions using CLAS12 standard background merging program \cite{Stepanyan:2020bg}. 

The background merging software uses real experimental data for given luminosity to extract background hits from all detector components that can later be overlayed on top of simulated data to emulate the realistic background conditions of the experiment. For our studies, we used background files from runs with beam currents of $45~nA$, $50~nA$, and $55~nA$. Combining them sequentially we generated data corresponding to $45~nA$, $95~nA$, and $150~nA$. The $95~nA$ data sample was produced by merging the $45~nA$ background file with the output of GEMC and then merging it with $50~nA$ background data. Similarly by merging $45~nA$, $50~nA$, and $55~nA$ in a sequence we obtained a data sample corresponding to $150~nA$. In further discussions, we refer to the original data sample simulated with Pythia and processed with GEMC without background as $0~nA$ data. All comparisons of single track efficiency and physics final state statistics 
are presented relative to those quantities obtained from the $0~nA$ data sample.

Most CLAS12 experiments so far have run with a $45~nA-50~nA$ beam on a liquid hydrogen target, and we want to measure the performance impact of the de-noising procedure for standard running conditions, and also see if we can run at higher beam currents (luminosity) which will increase the statistical power of experiments at given run time.

\subsection{Data Analysis}

To study the effect of the de-noising on particle reconstruction efficiency we processed the produced data samples through the stand-alone denoiser program to create de-noised counterparts of simulated data for each luminosity setting.  Both data samples were processed using the CLAS12 data reconstruction program. Then the track reconstruction efficiency was calculated for both data samples (original and de-noised) as a function of luminosity. The track reconstruction efficiency was calculated following the standard procedure for CLAS12~\cite{Stepanyan:2020bg}. The efficiency for positive tracks is defined as a ratio of 
events containing an electron and a positive hadron ($N_{eh^+}$) to the number of inclusive events with an electron reconstructed ($N_{e}$). The efficiency for negative tracks is calculated similarly:

\begin{equation}
L_t^+ = \frac{N_{h^+e}}{N_e} , L_t^- = \frac{N_{h^-e}}{N_e} 
\label{eq::eff}
\end{equation}
where $L_t^+$ is the multiplicity for positive particles and $L_t^-$ is the multiplicity for negatively
charged particles, respectively. In order to estimate the charged-particle reconstruction efficiency
as a function of the beam current, the multiplicity, $L_t^{+/-}$, is fitted with a linear function:
\begin{equation}
L_t^{+/-} = a + b\times I
\label{eq::eff2}
\end{equation}

Here $a$ and $c$ are the fit parameters and $I$ is the beam current. Then it is assumed that the
reconstruction efficiency, $E=1$ at $I=0$ nA:

\begin{equation}
E^{+/-} = 1 + c \times I
\label{eq::eff3}
\end{equation}
with $c=\frac{b}{a}$. The slope parameter $c$ represents the variation of the reconstruction
inefficiency per unit of the beam current ($nA$)~\cite{Stepanyan:2020bg}.

\subsection{Artificial Intelligence Assisted Tracking}

The CLAS12 data reconstruction software already contains neural networks helping to identify track candidates from combinations of clusters reconstructed in each of the super-layers of drift chambers~\cite{Gavalian:2020mlp}.
This network already provides a big improvement in the tracking efficiency compared to the conventional reconstruction algorithm. The impact on physics (depending on the number of particles in the reaction) is a $15\%-35\%$ increase in statistics. 
In the recently developed reconstruction software, the user can choose to use assistance from AI in identifying tracks or use purely the conventional algorithm to identify track candidates. In our studies, we first investigated the improvement of the de-noising algorithm by using the conventional algorithm to identify tracks. Then we extended these studies to include AI track identification when processing raw and de-noised data. By doing this we want to disentangle the performance improvements arising from de-noising and from AI assistance. 

\section{Pseudo-data analysis with de-noising}

In this section, we compare results from the analysis of the background merged MC data sample with files that were de-noised prior to running through CLAS12 reconstruction software. The comparison is done for data samples with different luminosities (namely $45~nA$, $95~nA$, and $150~nA$ electron beam incident on a $5cm$ long liquid hydrogen target). The data for the raw sample and the de-noised sample is processed with the same settings of CLAS12 reconstruction software and the tracks reconstructed in each sample are analyzed.

\subsection{Luminosity dependence}

\begin{figure}[!h]
\begin{center}
 \includegraphics[width=3.1in]{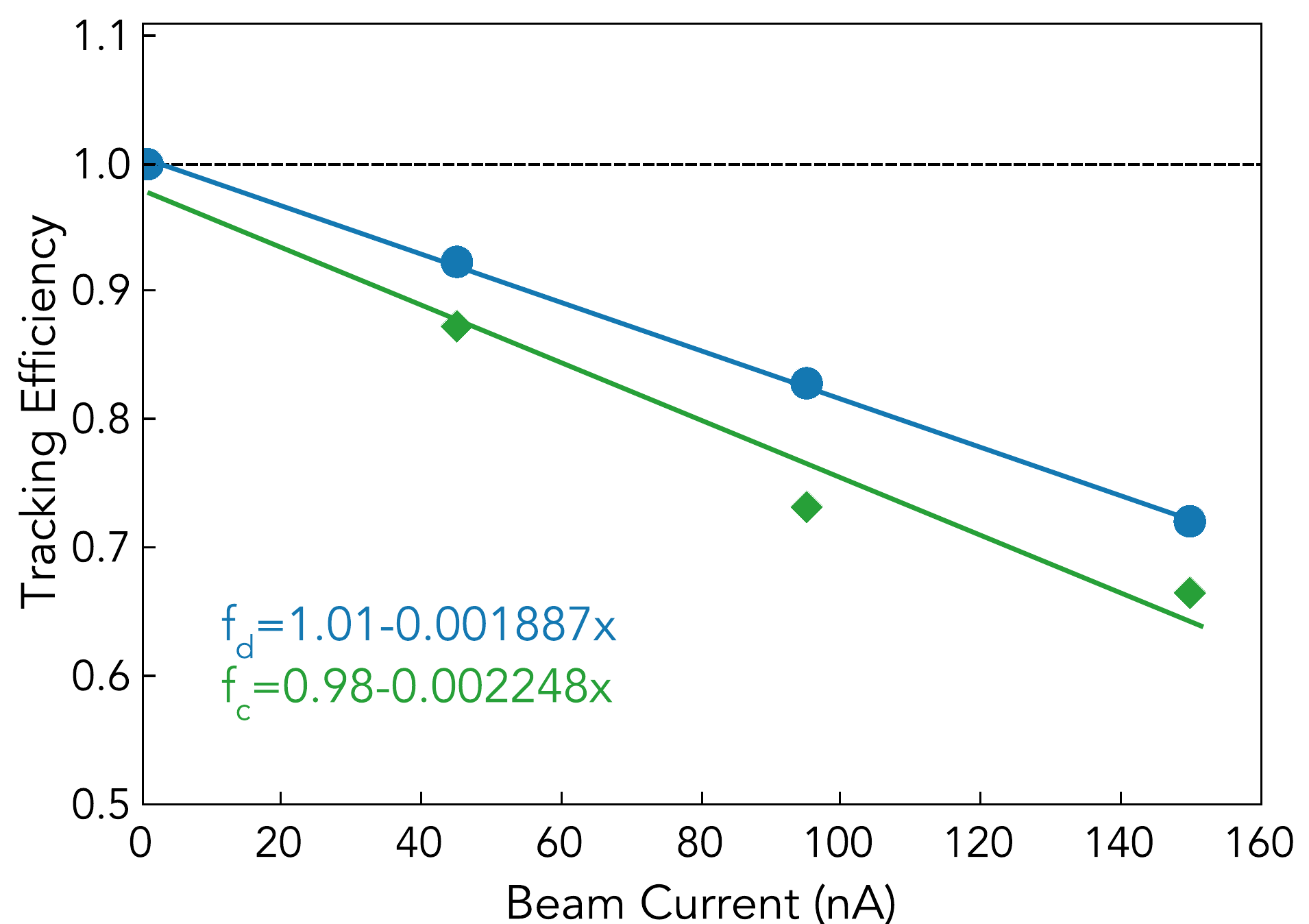}
 \includegraphics[width=3.1in]{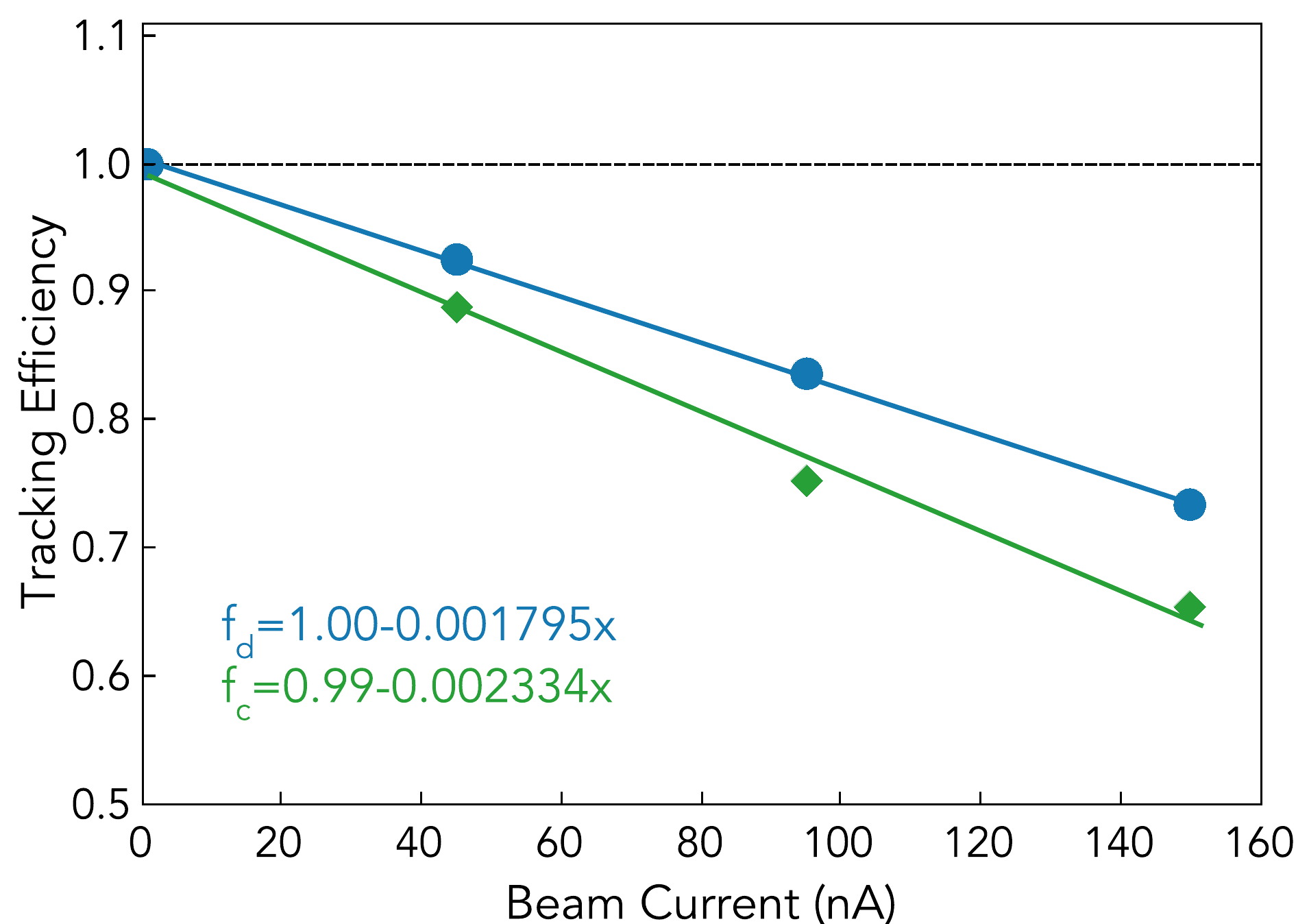}
\caption {Tracking efficiency as a function of luminosity (beam current) for positively (a) and negatively (b) charged particle.  The efficiency is shown for
conventional algorithm running on background merged files (diamonds), and on files with merged background then de-noised with AI (circles).}
 \label{lscan::conv_dn}
 \end{center}
\end{figure}

The track reconstruction efficiency is calculated according to Eqs.~(\ref{eq::eff}), (\ref{eq::eff2}) and (\ref{eq::eff3}) for positively and negatively charged particles. The results are shown in Figure~\ref{lscan::conv_dn}. The track reconstruction efficiency is an integrated quantity over the particle phase space. In our studies, we used a pre-selected simulation sample of three particles in the final state, which does not necessarily have angular and momentum dependence similar to experimental data and the efficiency dependence on beam current can reflect this. In these studies, we show a relative increase in efficiency when our methods are applied to simulated data.

As can be seen from the figure the number of reconstructed hadron-electron pairs relative to the number of reconstructed electrons is higher for the de-noised data sample compared to the raw data sample. This is due to an increased number of clusters reconstructed by the conventional clustering algorithm in the de-noised data samples. Detailed studies of cluster reconstruction efficiency are performed 
in our previously published article~\cite{Thomadakis:2022zcd}. 
The results show that the slope of the efficiency degradation as a function of the luminosity is significantly improved in the de-noised data sample. 
It is worth noting that the track reconstruction efficiency at 75 nA with a de-noised data sample is the same as for the 
$45~nA$ when reconstructing raw data sample (without de-noising). This implies that the experiment can run effectively at $75~nA$, collecting data 
twice faster while maintaining the same track reconstruction efficiency, which will lead to higher experimental significance in measured observables. 

\subsection{Physics Impact}

The processed data was also evaluated to extract physics observables from both data samples to discern the impact on physics for the de-noising algorithm. As mentioned before, the data selected from the Pythia simulation was for the final state $H(e,e^\prime\pi^+\pi^-p)X$ containing exactly four charged particles. From this sample the missing mass distribution of $H(e,e^\prime\pi^+\pi^-)X$ is analyzed showing a peak around proton mass where the selected reaction is inclusive $\rho$ meson production and some background (above proton mass) where other reactions are present (with missing neutral particles).

\begin{figure}[!h]
\begin{center}
 \includegraphics[height=3.1in]{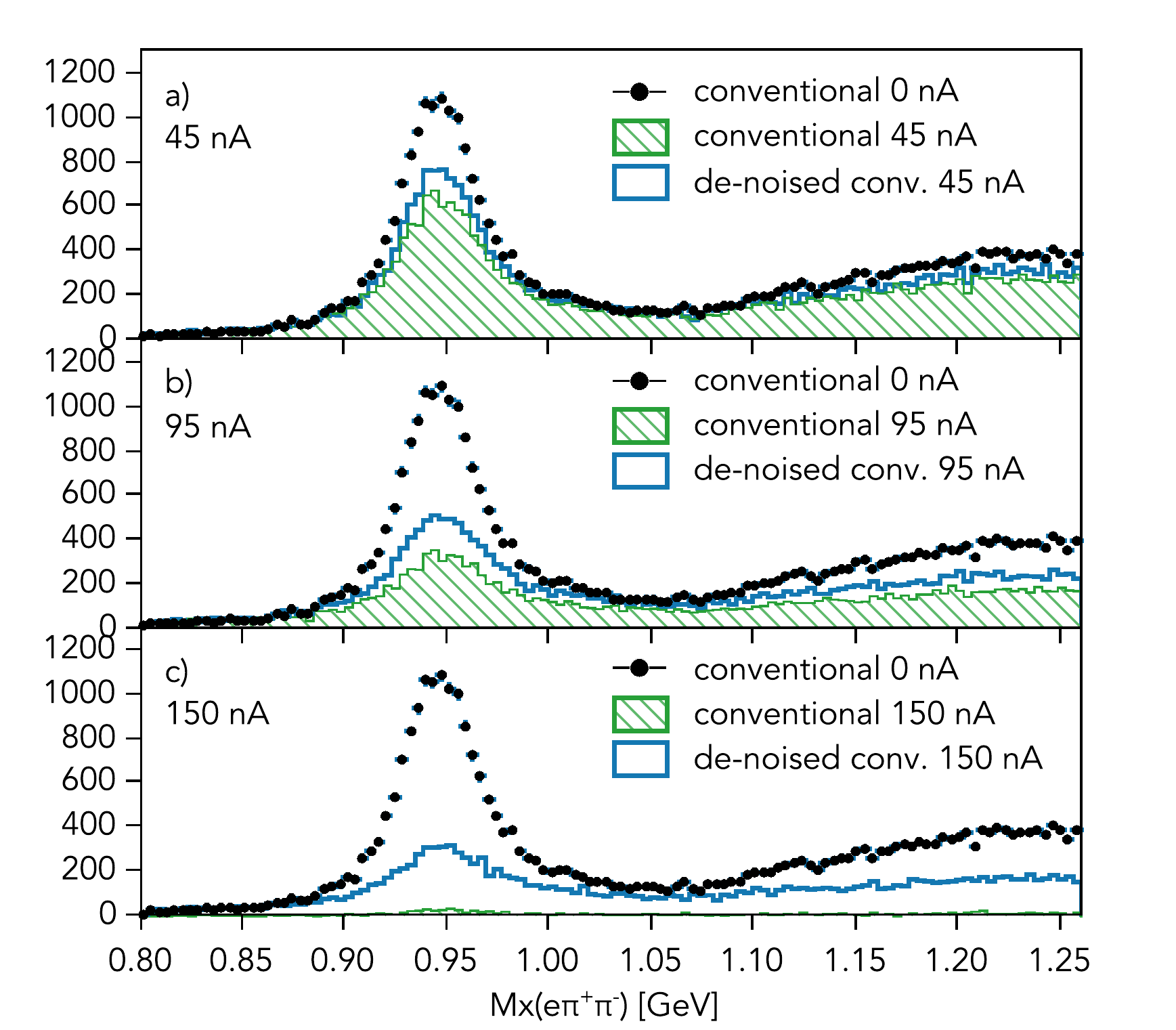}
   \includegraphics[height=3.1in]{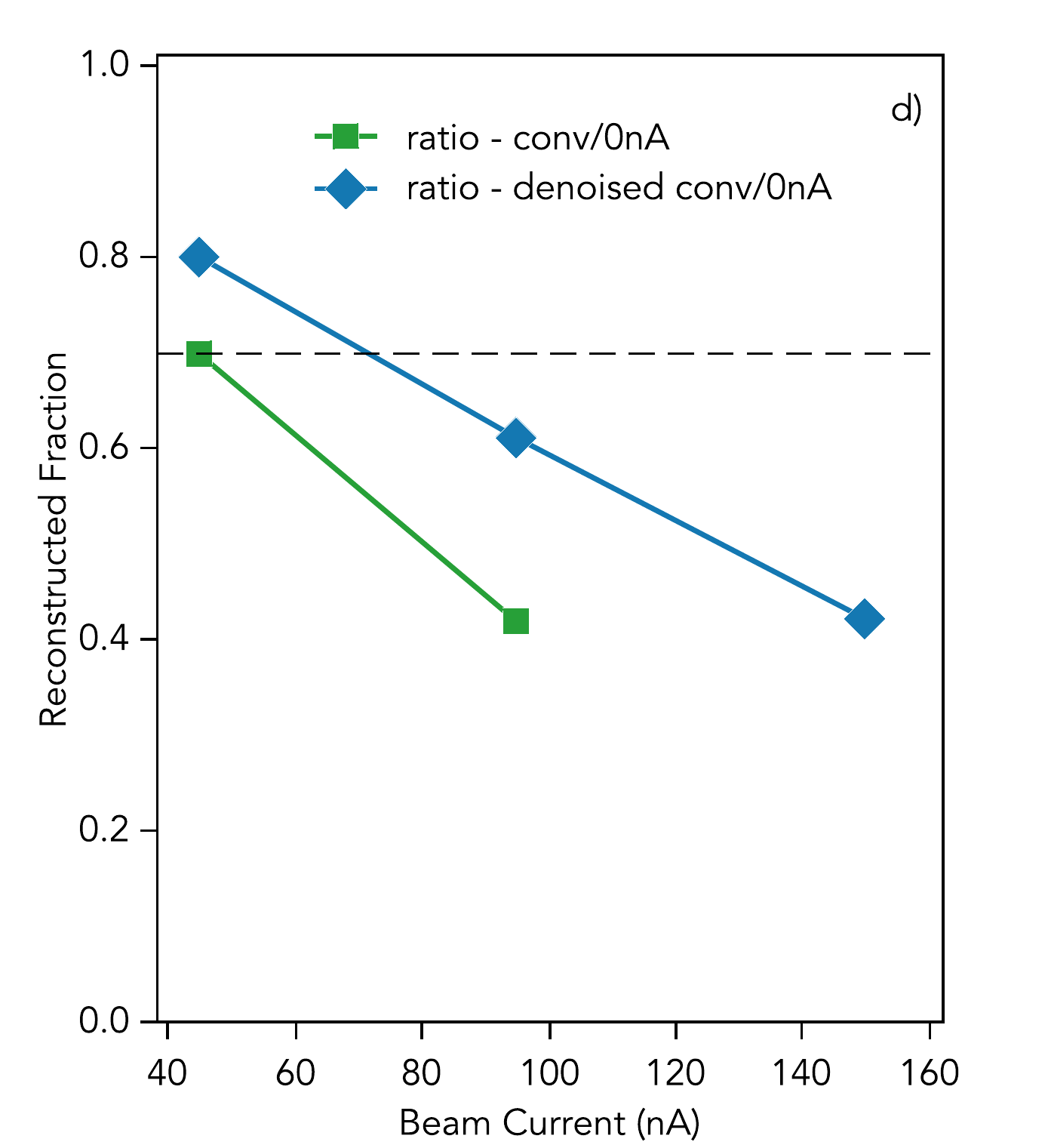}
\caption {
The de-noised data sample 
reconstructed with conventional algorithm (diamonds) for $45~nA$, $95~nA$ and $150~nA$. a), b) and c) reconstructed 
missing mass distributions for background merged data set reconstructed with conventional tracking (filled histogram) and
de-noised data sample reconstructed with conventional algorithm (solid line histogram).
d) The number of reconstructed protons from missing mass of $H(e \rightarrow e^\prime \pi^+\pi^-)X$ 
for background merged data set reconstructed with conventional tracking (squares) compared to de-noised data sample 
reconstructed with conventional algorithm (diamonds) for $45~nA$, $95~nA$ and $150~nA$.  }
 \label{physics::conv_dn}
 \end{center}
\end{figure}

In Figure~\ref{physics::conv_dn} the results of the analysis are shown, where the missing mass distribution $H(e,e^\prime\pi^+\pi^-)X$ is shown for different beam currents, in panels a), b) and c) the histograms show relative reconstructed distributions. The graph with points shows the missing mass reconstructed by the conventional tracking algorithm before any background is $0~nA$ for reference. The filled histogram shows the missing mass distribution reconstructed from background merged data with the conventional algorithm. The solid line histogram is the missing mass distribution reconstructed by a conventional algorithm after the background merged file is processed with a de-noising neural network to remove noise hits.
The summary of the number of protons in the missing mass distribution relative to the original (no background merged) distribution is presented in Figure~\ref{physics::conv_dn} d). It can be seen from the figure that the conventional algorithm reconstructs more tracks after de-noising the data. The number of reconstructed proton final states at $75nA$ from de-noised data is equal to the number of reconstructed final states at $45nA$ when using conventional track reconstruction algorithms.  Conducting experiments with higher incident beam current allows accumulating the necessary statistics for the proposed experiments in significantly less time, leading to huge savings in accelerator operations.

\section{Analysis of De-Noising data with AI assistance}

The two data samples, background merged and de-noised were also processed with the new 
reconstruction software, which includes AI-assisted track candidate identification~\cite{Gavalian:2020oxg},\cite{Gavalian:2020xmc}. 
The reconstruction software is designed to be able to process data in two parallel branches: in one 
branch it reconstructs tracks with the conventional algorithm where track candidates are identified by fitting all 
combinations of clusters forming a candidate and choosing candidates that pass the ``goodness'' of the fit criteria; 
 and in the second branch, AI classifies tracks from the list of candidates created from all combinations of clusters 
 forming a track. 
 The details on track candidate identification, software implementation, and the resulting outcome for increased 
 track reconstruction efficiency can be found in~\cite{Gavalian:2022hfa}.
 Two samples were processed and a comparison was made between conventional tracking algorithms from raw 
 background merged files and the output of the de-noised data sample with and without AI-assisted tracking. 

\subsection{Luminosity dependence}

\begin{figure}[!h]
\begin{center}
 \includegraphics[width=3.1in]{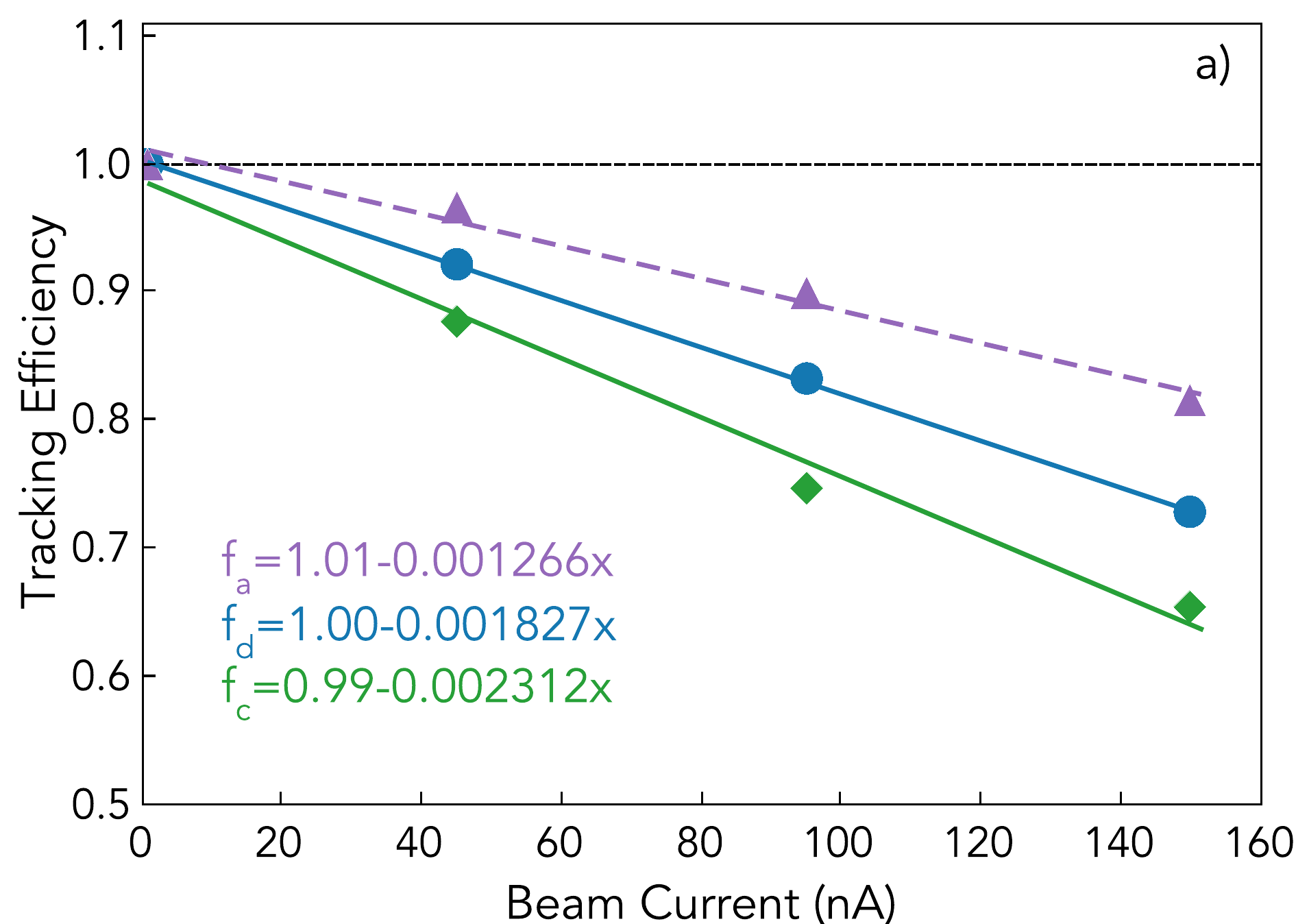}
 \includegraphics[width=3.1in]{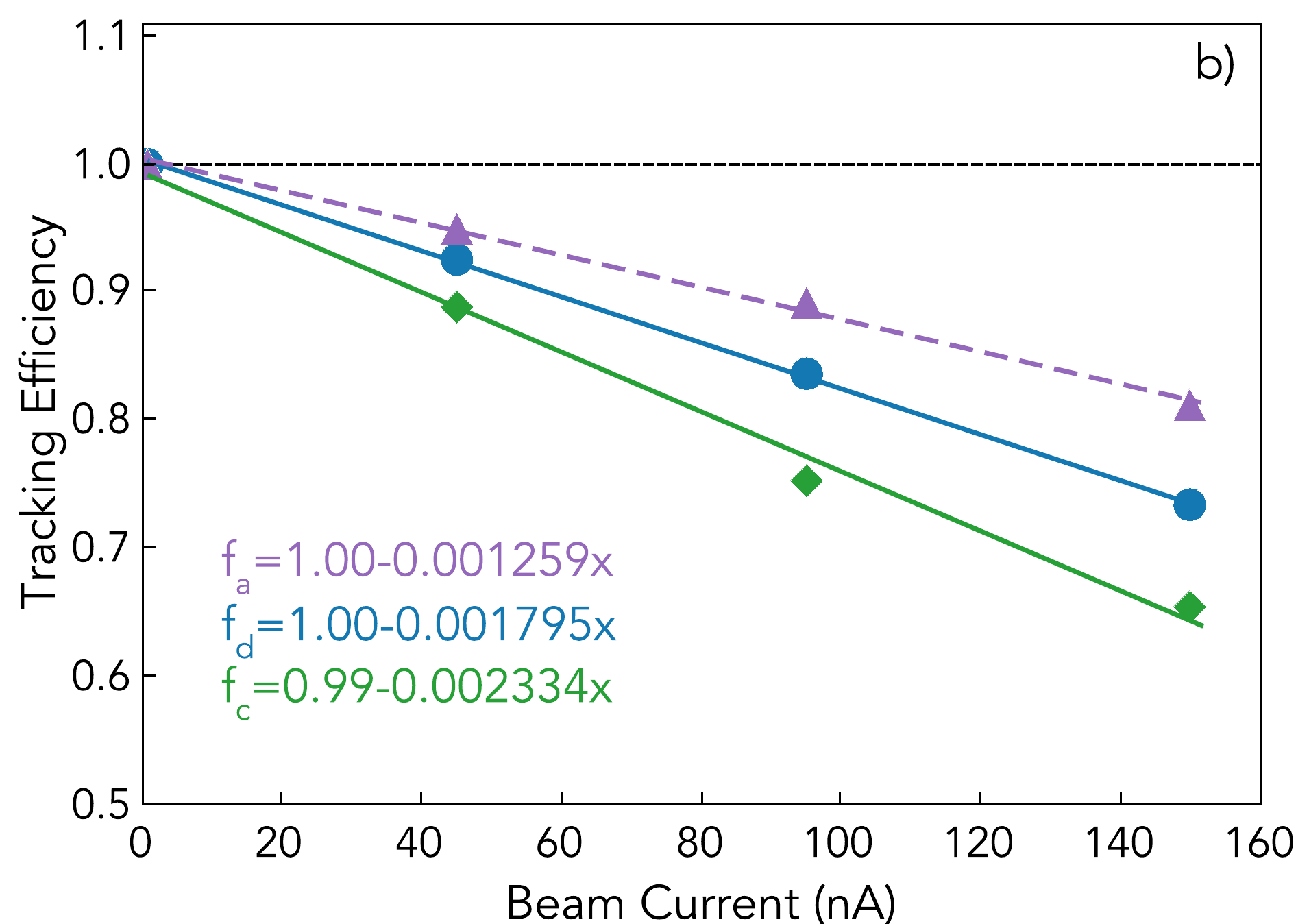}
\caption {Tracking efficiency as a function of luminosity (beam current) for positively (a) and negatively (b) charged particles.  The efficiency is shown for
conventional algorithm running on background merged files (diamonds), and on files with merged background then de-noised (circles), and de-noised data reconstructed 
 with AI assistance (triangles).}
 \label{lscan::conv_dn_ai}
 \end{center}
\end{figure}

The track reconstruction efficiency was calculated for the three samples using Eqs.~(\ref{eq::eff}),(\ref{eq::eff2}) and (\ref{eq::eff3}).
The results are presented in Figure~\ref{lscan::conv_dn_ai}. It can be seen from the figure that using 
AI-assisted tracking on the de-noised data sample further improves reconstruction efficiency. The raw background 
merged data sample exhibits a tracking efficiency decline of $0.23\%$ per nA, while the combination of de-noising 
and AI-assisted tracking reduces this slope to $0.12\%$ per nA (almost a factor of 2), resulting in an efficiency of $86\%$ at
 beam current $150~nA$ compared to $88\%$ at $45~nA$ beam current. 

This is a significant improvement in tracking efficiency when using both AI-assisted tracking with de-noising for a beam current 3 times higher than the current data collecting conditions. 

\subsection{Physics Impact}

Furthermore, the physics impact was studied for the de-noised data sample processed with AI-assisted tracking. The same data 
sample was used in these studies with selected $H(e,e^-\pi^+\pi^-p)X$ event from Pythia Monte-Carlo simulations, and analyzed for
missing mass of $H(e,e^-\pi^+\pi^-)X$, where the number of protons was extracted from the missing mass distribution.

\begin{figure}[!h]
\begin{center}
 \includegraphics[height=3.1in]{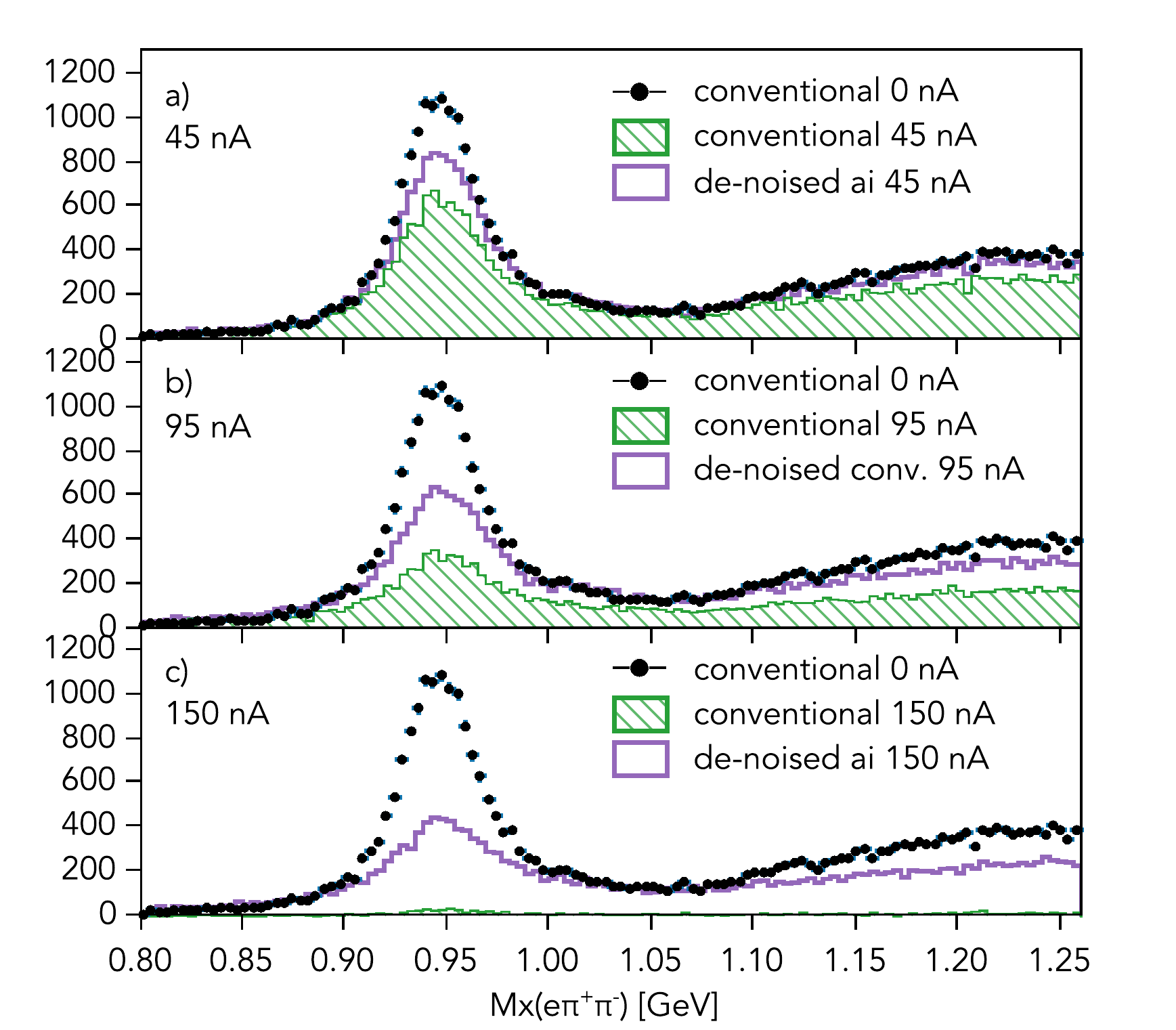}
 \includegraphics[height=3.1in]{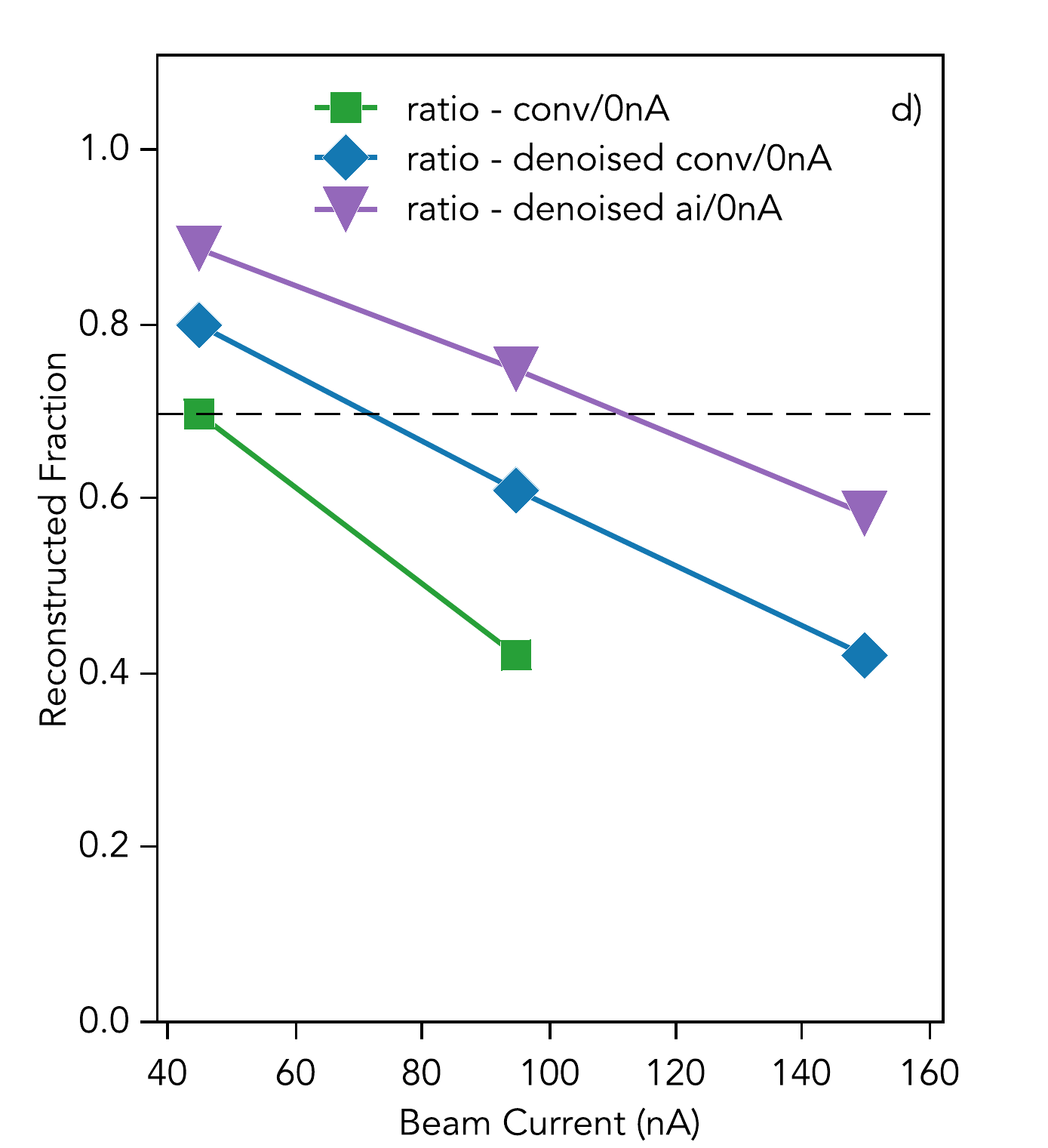}
\caption { 
The de-noised data sample was reconstructed with an AI-assisted tracking 
algorithm (triangles)  for $45~nA$, $95~nA$, and $150~nA$. a), b) and c) reconstructed missing mass distributions for 
background merged data set reconstructed with conventional tracking (filled histogram) and de-noised data sample 
reconstructed with AI-assisted algorithm (solid line histogram). Missing mass distribution for data sample before 
background merging ($0~nA$) is shown (circles) for reference.
The number of reconstructed protons from missing mass of $H(e \rightarrow e^\prime \pi^+ \pi^-) X$ 
for background merged data set reconstructed with conventional tracking (squares) compared to de-noised data sample 
reconstructed with conventional algorithm (diamonds) d). }

 \label{physics::conv_dn_ai}
 \end{center}
\end{figure}

The distributions of missing mass spectra are shown in Figure~\ref{physics::conv_dn_ai} for different beam current backgrounds.
In a), b) and c) the missing mass distributions are shown for the background merged data samples processed with the conventional algorithm 
(filled histogram) and the reconstructed missing mass after data de-noising and reconstructing with AI-assisted tracking (line histogram).
The graphs (circles) on all three plots show the missing mass distribution reconstructed from the generated data sample 
before any background is added for reference. In Figure~\ref{physics::conv_dn_ai} d) the summary of the studied data samples 
is presented. The background merged data samples analyzed with the conventional tracking algorithm (squares) show a sharp decline in
the number of reconstructed protons in the missing mass peak. Pre-processing data with the de-noising auto-encoders and processing
with the conventional algorithm (diamonds) improves the physics outcome due to improved single-track efficiency. The biggest improvement
comes from using AI-assisted track classification software after de-noising the drift chambers data (triangles). 


\section{Discussion}

Studies with simulated data indicate that using de-noising auto-encoders significantly improves the performance
of the conventional CLAS12 tracking algorithm, see Figure~\ref{physics::conv_dn} d). Further improvements come from using
the already established AI-assisted track classifier network with the de-noised data, see Figure~\ref{physics::conv_dn_ai} d).

It is evident from these studies that the analysis of existing data can benefit from this approach to tracking by an increase 
of statistical significance of physics observables. The numbers for reconstructed protons for each background setting 
and method of track reconstruction are summarized in Table~\ref{table:summary}. Using de-noising and AI-assisted 
tracking the statistics (in this particular case of three detected particles) increases by $26\%$. 

\begin{table}
\begin{center}
\begin{tabular}{l|ccc}
Stats & Conventional & De-noised & De-noised + AI CL \\
\hline
 nucleons (45 nA)  & 27225 &  30576 & 34277 \\
 nucleons (95 nA)  & 17125 & 23845 & 29428 \\
 nucleons (150 nA) &  1576 & 17018 & 23601 \\
\hline
\hline
ratio to conventional 45nA & 1.0 & 1.12 & 1.26 \\
ratio to conventional 95nA & 1.0 & 1.39 & 1.72 \\
ratio to conventional 150nA & 1.0 & 10.80 & 14.97 \\
\end{tabular}
\end{center}
\caption{Number of extracted nucleons from missing mass distribution for different beam currents
and different reconstruction methods. The bottom of the table presents the ratio of the number of nucleons for
different methods to the number for conventional tracking algorithm at $45~nA$ for all incident beam currents.}
 \label{table:summary}
\end{table}

As can be seen from the table, background merged simulated data processed with de-noising and the
 AI-assisted tracking leads to more events in the missing mass peak than data reconstructed with the 
 conventional tracking algorithm.
 
Conducting experiments with $95~nA$ incident beam energy will take twice less time to accumulate the same number 
of events as at 45 nA analyzed with the conventional tracking.

Even though the number of reconstructed nucleons is bigger when running at $45~nA$ and using improved 
tracking (including AI de-noising and the AI classifier), the argument can be made that the collected statistics at 
$95~nA$ (because of the rate of interactions at higher incident beam current) will lead to more physics relevant
statistics even with slightly lower track reconstruction efficiency.
The second half of Table~\ref{table:summary} shows the ratio of the number of nucleons in the 
missing mass peak for different beam currents and algorithms used. It can be seen that with increased 
beam current the denoiser gain over the conventional algorithm is exponentially increasing, indicating that 
the denoiser is very efficient in isolating hits that potentially belong to a ``true'' track candidate.
This study suggests that augmenting tracking algorithms with artificial intelligence opens the possibility
of conducting experiments at higher luminosity collecting larger data samples for physics reactions in 
a shorter time. This will definitely affect the estimation of experimental running conditions for the CLAS12 
detector for future experiments.

\section{Summary}

In this article we present a Machine Learning approach to de-noising detector data, the CLAS12 drift chambers specifically, using Convolutional Auto-Encoders. The data processed with the neural network and further processed with conventional tracking resulted in a significant increase in the number of reconstructed tracks. The study performed on simulated data shows a significant improvement in track reconstruction efficiency as a function of experimental luminosity. Using de-noising in combination with AI-assisted tracking further improves the track reconstruction efficiency. 
The resulting increase of physics events in MC is estimated to be $26\%$ for three-particle final Staten the reaction 
$H(e,e’\pi^+\pi^-)p$ for the nominal experimental luminosity of CLAS12 ($45~nA$ electron beam and a $5~cm$ long 
liquid hydrogen target). The efficiency of track reconstruction from the $95~nA$ beam background merged MC data using the 
de-noising and AI-assisted tracking are equal to the efficiency of the conventional tracking from the $45~nA$ background merged data.
This can lead to the possibility of running experiments at a higher luminosity and accumulating the same physics statistics in $2.5$ times shorter time.

\section{Acknowledgments}

This material is based upon work supported by the U.S. Department of Energy, Office of Science,
Office of Nuclear Physics under contract DE-AC05-06OR23177, and NSF grant no. CCF-1439079 and
the Richard T. Cheng Endowment. This work was performed using the Turing and Wahab computing
clusters at Old Dominion University.
 
\newpage
\bibliography{references}
\bibliographystyle{ieeetr}

\end{document}